\documentclass[reprint,amsmath,amssymb,aps,prl,noeprint,nolongbibliography]{revtex4-2}
\usepackage[utf8]{inputenc}
\usepackage{graphicx}
\usepackage{amsmath}
\usepackage{xspace}
\usepackage{chemformula}

\begin{document}

\title{Atomic structure optimization with machine-learning enabled interpolation between chemical elements}
\author{Sami Kaappa}
\affiliation{Department of Physics, Technical University of Denmark, Kongens Lyngby, Denmark}

\author{Casper Larsen}
\affiliation{Department of Physics, Technical University of Denmark, Kongens Lyngby, Denmark}

\author{Karsten Wedel Jacobsen}
\affiliation{Department of Physics, Technical University of Denmark, Kongens Lyngby, Denmark}
\email{kwj@fysik.dtu.dk}
\date{\today}

\newlength{\figwidth}
\setlength{\figwidth}{0.95\columnwidth}
\newlength{\widefig}
\setlength{\widefig}{0.9\textwidth}

\begin{abstract}
We introduce a computational method for global optimization of structure and ordering in atomic systems. The method relies on interpolation between chemical elements, which is incorporated in a machine learning structural fingerprint. The method is based on Bayesian optimization with Gaussian processes and is applied to the global optimization of Au-Cu bulk systems, Cu-Ni surfaces with CO adsorption, and Cu-Ni clusters. The method consistently identifies low-energy structures, which are likely to be the global minima of the energy. For the investigated systems with 23-66 atoms, the number of required energy and force calculations is in the range 3-75.
\end{abstract}

\maketitle

The atomic-level structure of a material is often of crucial importance for its mechanical, electronic, magnetic, or chemical properties. At low temperatures, the structure can in principle be determined computationally by minimizing the energy of the system. However, the space of different atomic configurations of a material is huge, and structure determination is therefore a longstanding challenge in computational physics.

A number of methods to address this challenge have been introduced \cite{Zhang2020} including basin hopping \cite{basinhopping}, particle swarm optimization \cite{SGO}, evolutionary algorithms, \cite{EA2, sioclusters, Lysgaard2014, Jaeger2019},  and random searches \cite{randomsearch}. However, all of these methods require a large number of energy and force calculations, which can be time consuming if performed with for example density functional theory (DFT) or higher-level quantum chemistry methods.

During the last decade, machine learning techniques have gained impact on computational material physics \cite{Schmidt2019, rinke2019, Butler2018, Bartok2017, Behler2016, denzel2020hessian, Huan2017, GPMin, bondmin, Deringer2017machine, GarridoPRL2019, koinstinen2017, Chmiela2016FF, schnet, burke2021, BEEFvdw, ceriottidensity,GPMin}. The field of global optimization of atomic structures took a major step forward recently with the GOFEE approach \cite{Malthe2020efficient}. In this approach Gaussian processes are used to generate a surrogate potential energy surface (PES), which is then explored by random searching and Bayesian optimization. Thereby, a speedup of several orders of magnitude in determining the optimal structure are in some cases achieved \cite{Malthe2020efficient}. This method was expanded with training to forces and several other modifications in the BEACON code \cite{beacon}.

Global optimization is particularly challenging if many local minima exist in the PES. This would typically be the case for large systems, but it also appears for smaller systems when several chemical elements are present. In many alloys, the interchange of two atoms introduces another local minimum in the potential energy surface but with a different energy. The number of local minima thus becomes the number of different ordering permutations of the atoms in the lattice, which can be huge. For example, in a 64-atom unit cell with equal amounts of two different types of atoms, the number of possible permutations is of the order of 10$^{18}$, and it is of course not possible for any algorithm to explore all these configurations.

Here we address the challenge of many local minima by interpolation between chemical elements (``ICE''). With this approach atoms of different chemical elements can switch place not by moving in real space but by gradually changing their chemical identities. This leads to a much reduced or even vanishing barrier for the process. The idea of interpolating in chemical space has also been used in other contexts including catalyst design \cite{Jacobsen:2001kl}, in combination with perturbation theory \cite{Weigend.2014,PhysRevLett.95.153002,Rudorff:2020, huang2020ab}, or to treat disordered alloys with the coherent potential approximation \cite{Soven:1967fd,Ruban.1994,Abrikosov.1994}. We integrate the idea of interpolation between chemical elements with BEACON into a new method (``ICE-BEACON''), which allows for global structure optimization through simultaneous optimization of chemical identities and atomic coordinates.

At the heart of the method is a fingerprint, which is a vector representation of the system configuration for a given set of atomic coordinates and fractional chemical identities of the atoms (termed ``element fractions'' in the following). For all element fractions equal to 0 or 1, the fingerprint is the same as the one used in BEACON \cite{beacon}, and it is described in detail in the Appendix. Given a database of energy and force calculations obtained for example with DFT, the fingerprint allows the construction of a surrogate PES using a Gaussian process. It is possible to analytically predict energies and the derivatives with respect to both atomic coordinates and element fractions on the surrogate PES, allowing the use of gradient-based optimization algorithms.

\begin{figure*}[ht]
    \centering
    \includegraphics[width=\widefig]{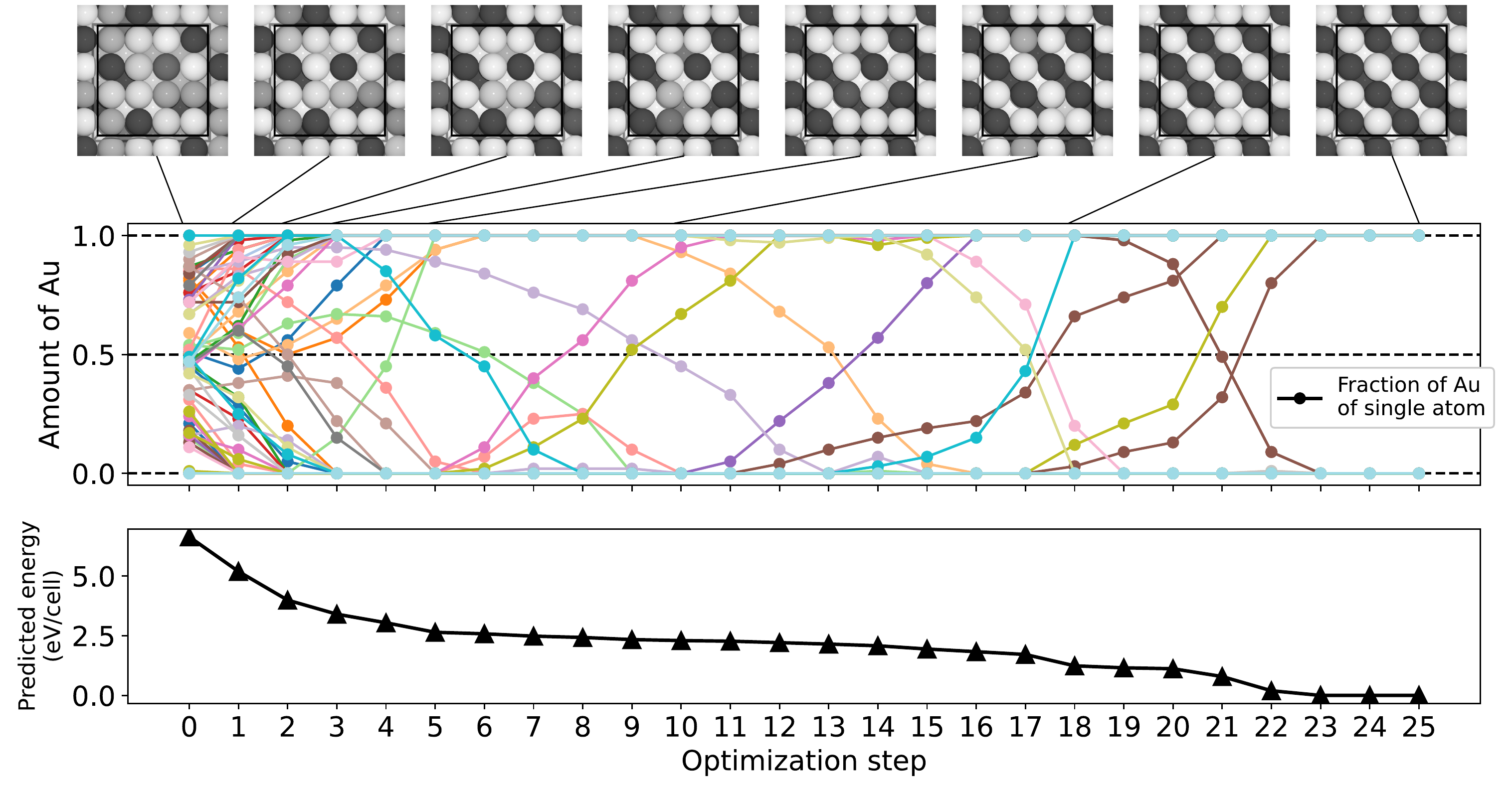}
    \caption{A single local optimization run of element fractions in AuCu bulk with 64 atoms in a unit cell. The training set consists of 4 DFT calculations. The structures at different steps are shown above with the color saturation denoting the corresponding element fractions (black: Au and white: Cu). The black lines indicate the computational supercell boundaries. In the middle panel, the evolution of the element fractions (where Au corresponds to 1 and Cu to 0) is shown for all atoms along the optimization. In the lower panel, the energy predicted by the surrogate model is shown. The optimization is seen to proceed to the ground state without getting trapped in metastable states.}
    \label{fig:param_opt}
\end{figure*}

The global optimization uses Bayesian optimization and is performed in the following way. (1) An initial database of typically two DFT calculations based on random configurations is constructed. (2) The associated surrogate PES is explored by 40 energy minimizations starting from random configurations in the space of element fractions and atomic coordinates, or from previously visited energy minima with rattling of the atomic positions and with random element fractions. The element fractions are constrained to keep the right overall stoichiometry. (3) The element fractions of the resulting 40 configuration are rounded to the nearest integer (if necessary), and the configurations are evaluated with a lower-confidence-bound acquisition function (see Appendix for details). A DFT calculation for the configuration with the lowest acquisition function value is included in the database. The procedure is continued by returning to step (2).

All the DFT calculations presented in this paper are performed with GPAW \cite{gpaw} using the Atomic Simulation Environment \cite{ase,ase-paper} and the PBE functional \cite{PBE} unless otherwise stated. The local relaxations in the surrogate potential energy surface are carried out by sequential least squares programming (SLSQP) as implemented in the SciPy package \cite{slsqp, scipy} that enables the use of both equality and inequality constraints with efficient gradient-driven optimization. In ICE-BEACON, we need equality constraints to fix the number of atoms of different elements in the system, while inequality constraints are used in order to limit the element fractions to be between 0 and 1.

We first consider a AuCu bulk system in a fixed fcc structure with a 64 atom supercell with 32 atoms of both gold and copper, and a lattice parameter of 3.767 \AA. In Fig.~\ref{fig:param_opt}, we illustrate a single local optimization of the atomic ordering by relaxing the element fractions of the atoms within the surrogate potential energy surface. The training set consists of 4 DFT calculations with different orderings of the atoms.

Starting from random fractions (step index 0), the run ends up with the correct layered structure of Au and Cu \cite{Okamoto1987, AuCu-Kubiak}. The energy evolution is smooth and the descent optimization is able to guide the system into the low-energy minimum without trapping in states with non-integer element fractions. 

We also see that some of the parameters stay at 0 or 1 during multiple steps, but then change from 0 to 1 (or from 1 to 0) later in the run. This shows that the method not only pushes all the fractions to the closest zeros and ones (which would be close to a local minimum in coordinate space), but it is able to circumvent energy barriers by interpolating in chemical element space.

\begin{figure}[ht]
    \centering
    \includegraphics[width=\figwidth]{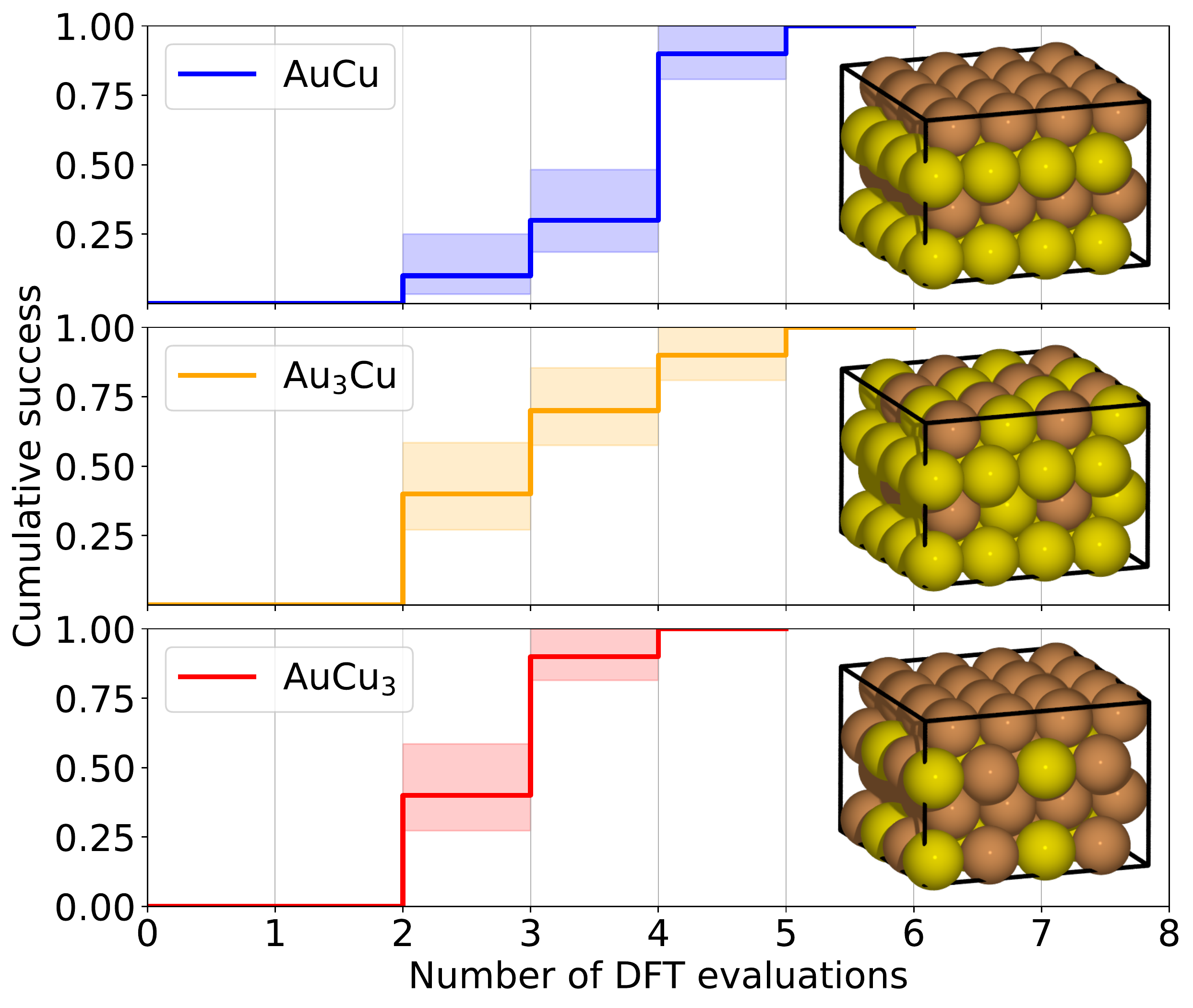}
    \caption{Success curves for optimizing the ordering of \ch{Au_x Cu_y} in a bulk fcc lattice using ICE-BEACON. For each system, 10 individual optimization runs are performed. The number of DFT evaluations in the training set is shown on the $x$-axis, while the fraction of successful runs, where the correct structure is identified, is shown on the $y$-axis. The indicated uncertainties are calculated with bootstrapping.}
    \label{fig:aucu_sc}
\end{figure}

We now consider global ordering optimization of the AuCu, \ch{Au3Cu}, and \ch{AuCu3} bulk systems. The total number of atoms for each supercell is 64 (see Fig.~\ref{fig:aucu_sc}). Even though all atoms are fixed at lattice positions, we train the surrogate PES also on forces and thereby include additional information about the local environment around the atoms. Fig.~\ref{fig:aucu_sc}  shows the success curves for each system. The essential features of the PES are learned very quickly, as we only need at maximum 5 DFT evaluations to obtain the correct ordering in all of the 10 runs per system. The identified structures match the experimentally determined ones at low temperatures \cite{Okamoto1987}. Experimentally, AuCu exhibits an interlayer expansion between the Au and Cu layers breaking the cubic symmetry \cite{Okamoto1987, AuCu-Kubiak}, but including the distortion does not change the result as discussed further in the Supplemental Material \cite{suppinfo}.

We note that the algorithm does not know when the global minimum ordering is found, and it thus continues the search after the eight DFT evaluations shown in the figure. It should be emphasized that the algorithm itself determines which configurations to evaluate with DFT based on the acquisition function. This ensures a balance between relevant low-energy structures and exploration of new configurations with large uncertainties.

To further investigate the applicability of ICE-BEACON, we study a CuNi alloy with an fcc(111) surface. The system is setup as a 4-layer slab with a $4\times 4$ surface supercell and a fixed in-plane lattice constant of 3.583 \AA~(64 atoms in total). We consider the situation with 16 Cu atoms. The coordinates and chemical identities of the bottom-most Ni layer are fixed during the optimizations. The GPAW calculations are run in spin-paired mode, but we have verified that spin polarization has no effect on the order of the energies for the final structures. The RPBE functional \cite{RPBE} is used for the {C}u{N}i surface systems.

\begin{figure}[t]
    \centering
    \includegraphics[width=\figwidth]{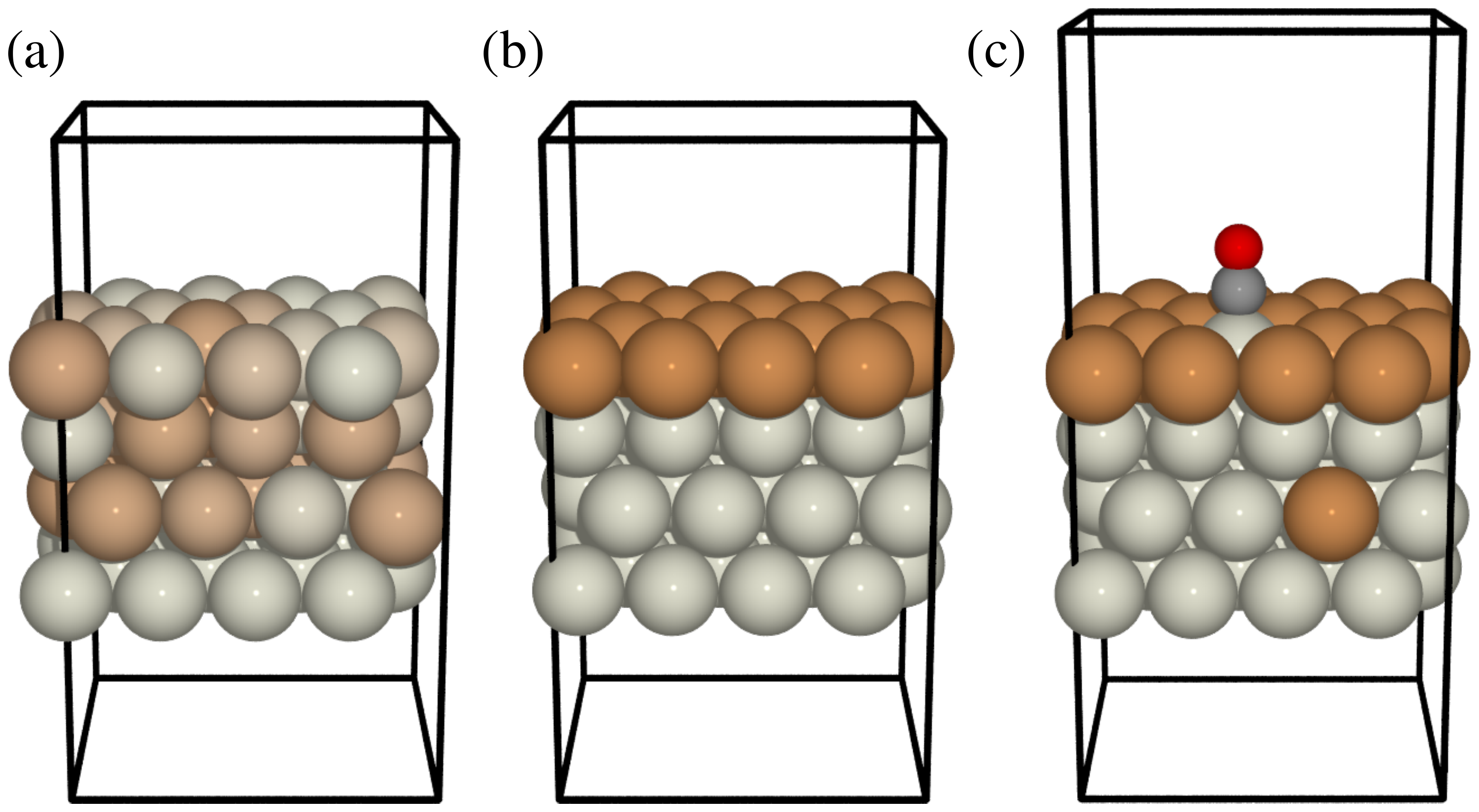}
    \caption{Initial and optimal CuNi surface structures with 16 Cu atoms out of a total of 64. (a) Typical initial structure before optimization of the clean surface. The colors are linearly scaled between those of Ni and Cu corresponding to the element fractions of the atoms. (b) Optimal structure for the clean surface system as found with ICE-BEACON with simultaneous optimization of coordinates and fractions. The structure is found after 3 DFT calculations in all of 4 runs. (c) Optimal structure after introducing a CO molecule at the surface. We see that the CO molecule pulls a Ni atom to the surface due to the stronger chemisorption. During the optimizations, the bottom-most layer is fixed to be Ni in fixed positions. Colors: Cu: brown, Ni: light grey, C: dark grey, O: red.}
    \label{fig:surfaces}
\end{figure}

The initial structures for the optimizations in the surrogate potential energy surfaces are obtained by rattling the atomic coordinates by 0.05 \AA\ around the lattice sites and assigning random element fractions to the atoms in the three top layers keeping the total number of Cu atoms at 16. One example of an initial structure is shown in Fig.~\ref{fig:surfaces}a.

The result of an ICE-BEACON optimization with simultaneous optimization of the atomic positions and the element fractions is shown in Fig.~\ref{fig:surfaces}b. This structure was proposed as the global minimum by all of 4 individual optimization runs after 3 DFT calculations. The copper atoms are seen to migrate to the surface, which is also what is found experimentally \cite{Webber_1986}.

In Fig.~\ref{fig:surfaces}c we show the result of an ICE-BEACON calculation where a CO molecule is placed on top of the CuNi(111) surface. Again both atomic coordinates and element fractions are optimized. A nickel atom is pulled to the surface with the CO in an on-top position. This is in agreement with the fact that the adsorption energy is larger on Ni(111) than on Cu(111)  \cite{C9CP00881K}, and this overcomes the segregation energy of copper. Experimentally CO is observed to occupy top sites in Cu-Ni(111) alloy surfaces with mixing of the metal atoms \cite{doi:10.1063/1.456185} even though CO prefers hollow sites on pure Ni(111) and Cu(111) \cite{C9CP00881K}. The effect of Ni atoms back-segregating to the surface exposed to CO has also been observed experimentally \cite{PhysRevLett.46.1529}.

The on-top site of CO is observed in on the average 20 DFT calculations in 5 out of a total of 8 optimization runs. One of the other runs suggests a bridge site, and one run suggests a hollow hcp adsorption site. The last run does not find a low-energy structure. In all cases, we find that the CO molecule lifts Ni atoms to the neighboring sites. See Fig.~S2 in the Supplemental Material \cite{suppinfo} for the structures and their energies.

\begin{figure}[ht]
    \centering
    \includegraphics[width=\figwidth]{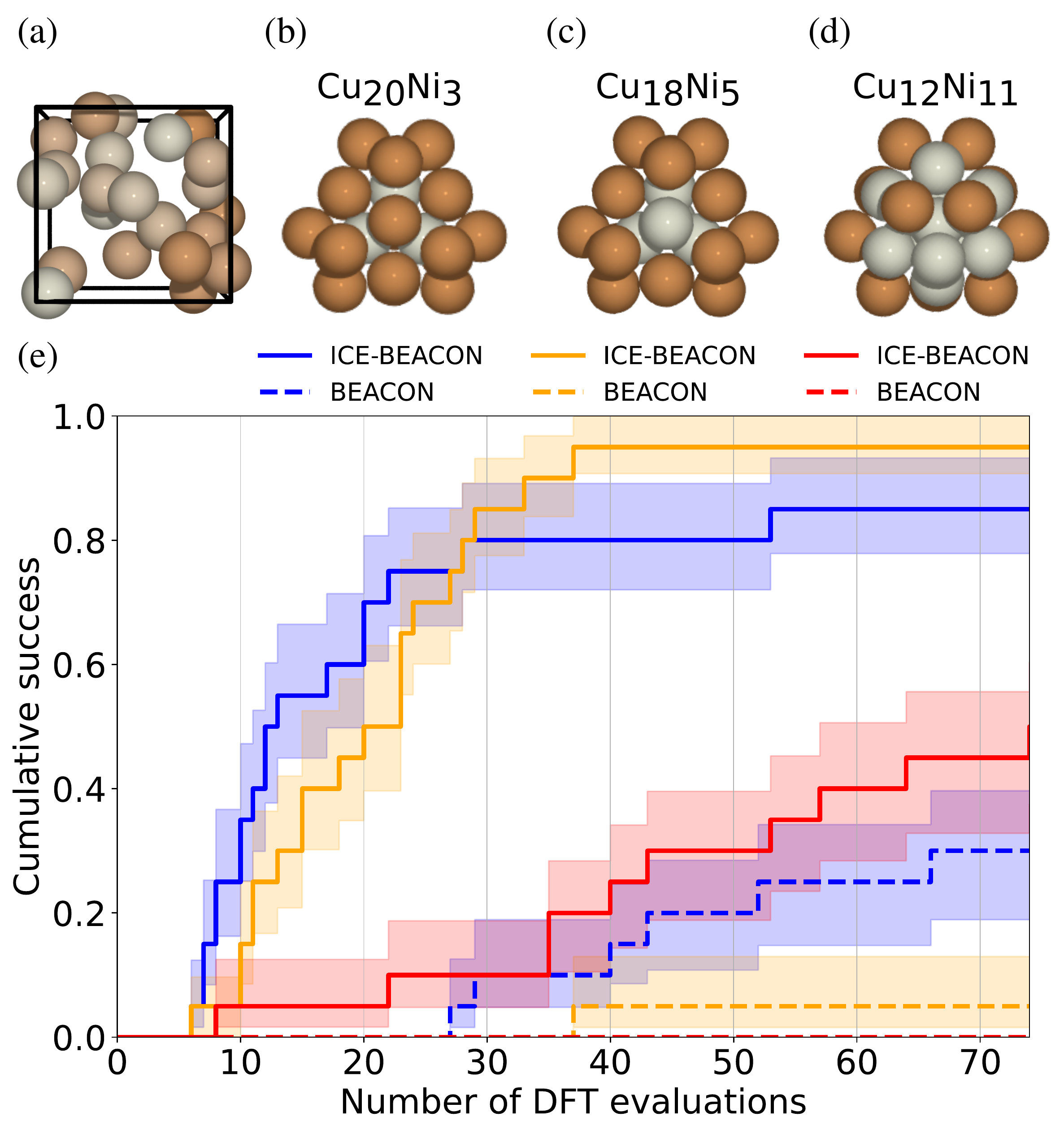}
    \caption{(a) Example of a random CuNi initial structure of a 23 atom cluster generated in a cubic box (black lines) before optimization in ICE-BEACON. The colors are linearly scaled between those of Ni and Cu corresponding to the element fractions of the atoms. (b), (c), and (d) The identified global minimum structures for \ch{Cu20Ni3}, \ch{Cu18Ni5}, and \ch{Cu12Ni11}, respectively, as determined with ICE-BEACON. (e) Success curves for optimizing the clusters with BEACON and ICE-BEACON up to 75 DFT calculations based on 20 runs each. ICE-BEACON requires considerably less DFT calculations than BEACON, and in the case of 11 nickel atoms, where the number of possible orderings exceeds a million only ICE-BEACON finds the optimal structure.
    }
    \label{fig:clusters}
\end{figure}

As the most challenging example, we consider CuNi clusters, where the metal atoms are not ascribed to lattice sites, and both element fractions and atomic coordinates are simultaneously optimized.  We consider CuNi clusters with a total of 23 atoms and varying content of nickel (3, 5, and 11 atoms). In Fig.~\ref{fig:clusters}, we compare the results of ICE-BEACON with BEACON and find a considerable reduction in the number of DFT calculations necessary to find the optimal structure when using ICE-BEACON.

For each cluster stoichiometry 20 ICE-BEACON optimization runs are performed and success is declared at the first encounter of an energy within 0.05 eV of the minimum energy encountered across all BEACON and ICE-BEACON runs for the respective cluster stoichiometry. Subsequently, it is checked that the structures within this energy range are in fact identical. The initial structures for the random searches in the surrogate PES are chosen with random initial atomic coordinates within a box of size $8\times 8\times 8\  \textrm{\AA}^3$, and with random element fractions (see Fig.~\ref{fig:clusters}a for an example). The computational unit cell is a cube with side length 16~\AA.

Figures \ref{fig:clusters}b-d show the identified structures. For 3 and 5 nickel atoms the structure is a 
triple-icosahedron, where 6 atoms are shared by two of the icosahedra and 5 atoms by all three. The latter are exactly the five nickel atoms in \ch{Cu18Ni5}. The cluster with 11 nickel atoms exhibits an icosahedron of 11 nickel atoms and 2 copper atoms with the icosahedron covered by additional copper atoms (in the back in the figure). The structure with 3 nickel atoms agrees with previous calculations based on an interatomic potential \cite{Hristova.2008}.

In the cases with 3 and 5 nickel atoms ICE-BEACON finds the optimal structures with typically 20 DFT calculations, while BEACON requires considerably more and does often not find the correct structure within 75 DFT calculations (see Fig.~\ref{fig:clusters}e). With 11 nickel atoms the number of possible different Cu/Ni decorations for given atomic positions becomes larger than a million and BEACON does not identify the optimal structure in any of the 20 runs. ICE-BEACON finds the structure in half of the runs with 75 DFT calculations.

To summarize, we have introduced a method to simultaneously optimize atomic positions and chemical ordering in atomic-scale systems. The method was illustrated with applications to bulk, surface, and cluster systems. The introduction of element fractions enables a significant improvement for multielement systems compared to a method like BEACON, which only works in coordinate space.

A striking feature of the method, and a key to its performance is that the interpolation between chemical elements allows for the switching of atoms of different types without an energy barrier. As a result the element fractions are not trapped in many metastable states during optimization. Another remarkable feature is that in most cases the final values for the element fractions are close to 0 or 1 as for real atomic materials. These features are present for all applications shown here, but to which extent this holds for other applications further research will show.

The approach can easily be extended to systems with several different chemical elements and further efficiency may be obtained by training the model to smaller subsystems as well. For systems where several different concentrations are investigated the surrogate model may benefit from a common database with all the DFT calculations.

The code for ICE-BEACON is included in the GPAtom package \cite{gpatom}.

\begin{acknowledgments}
We acknowledge support from the VILLUM Center for
Science of Sustainable Fuels and Chemicals, which is funded by
the VILLUM Fonden research grant (9455).
\end{acknowledgments}

\section{Appendix}
The fingerprint is the same that we use in BEACON with the essential difference that now the atoms are represented as fractional: each atom possesses the element fraction $q_A$ of one of the two elements and the fraction $q_B = 1-q_A$ of the other element. We use 
\begin{equation}\label{eq:fp_radial}
 { \rho}_{AB}^R(r, q; {\mathbf x}) = \sum_{\substack{i\in A \\ j\in B}} q_{i,A}q_{j,B}\frac{1}{r_{ij}^2}f_c(r_{ij}; R_c^R) \, e^{-|r-r_{ij}|^2/2\delta_R^2}
\end{equation}
for a radial distribution function and
\begin{align}\label{eq:fp_angular}
    \nonumber { \rho}_{ABC}^\alpha(\theta, q; \mathbf x) = \sum_{\substack{ i\in A \\ j\in B \\k\in C}} & \Big( q_{i,A} q_{j,B} q_{k,C} f_c(r_{ij}; R_c^\alpha)f_c(r_{jk}; R_c^\alpha) \cdot \\
    & e^{-|\theta-\theta_{ijk}|^2/2\delta_\alpha^2}\Big)
\end{align}
for an angular distribution, where $q_{i,A} \in [0, 1]$ is the element fraction describing how much atom $i$ is of element $A$. In Eqs.~\ref{eq:fp_radial} and \ref{eq:fp_angular}, $r_{ij}$ is the distance between atoms $i$ and $j$, and $f_c$ is a smooth cutoff function with cutoff distances $R_c$. The cutoff distances have $R_c^R = 8$ \AA~and $R_c^\alpha = 4$ \AA. The values for $\delta_R=0.4$ \AA~and $\delta_R=0.4$ rad are also kept fixed. The full fingerprint is obtained by concatenating the vectors created with Eqs. \ref{eq:fp_radial} and \ref{eq:fp_angular} for different element pairs and triples.

Energies and forces, $\mu=(E, -F)$, are calculated with the usual expression for a Gaussian process \cite{williams2006gaussian, poloczek2017gradients}
\begin{equation}\label{eq:predict}
    \mu(\mathbf x, Q) = \mu_{p}(\mathbf x) + K(\rho(\mathbf x, Q), P)C(P, P)^{-1}(y - \mu_p(X)),
\end{equation}
where $\mathbf x$ is the atomic coordinates, $Q$ are the element fractions, $\mu_p(\mathbf x)$ is the prior function, $K$ and $C$ are covariance matrices, $P$ is a matrix containing the fingerprints for the training data, and $y$ is the training targets. In Eq. \ref{eq:predict}, the covariance matrix $K$ is the only quantity that contains the information about the element fractions that are stored in the vector $Q$ for each atom. For details about the construction of the matrices $K$ and $C$ and the vector $y$, we refer the reader to the article about BEACON \cite{beacon}.  We use a constant prior $\mu_p(\mathbf x) = \mu_p$ and a squared-exponential kernel with a distance measure given by the Euclidean distance between the fingerprint vectors. Along an ICE-BEACON run, the prior constant and the kernel hyperparameters are updated by maximizing the marginal log-likelihood as also described in Ref.~\cite{beacon}.

Within the Bayesian optimization we use an acquisition function $f$ of the form of the lower confidence bound (LCB), that reads
\begin{equation}\label{eq:acq}
    f(\mathbf x) = E(\mathbf x) - 2\Sigma(\mathbf x),
\end{equation}
where $\Sigma(\mathbf x)$ is the uncertainty of the predicted energy, provided by the Gaussian process \cite{beacon, williams2006gaussian, poloczek2017gradients}.

\bibliography{bibliography}

\end{document}


\centering
{\large\bf{Supplemental Material for:\\ Atomic structure optimization with machine-learning enabled interpolation between chemical elements}}\\[10pt]

Sami Kaappa, Casper Larsen, Karsten Wedel Jacobsen\\

{\em{Department of Physics, Technical University of Denmark, Kongens Lyngby, Denmark}}\\

(Dated: \today)\\[30pt]
\flushleft


~\\[20pt]
{\bf{AuCu in elongated fcc lattice}}

\begin{figure}[h]
    \centering
    \includegraphics[width=0.6\linewidth]{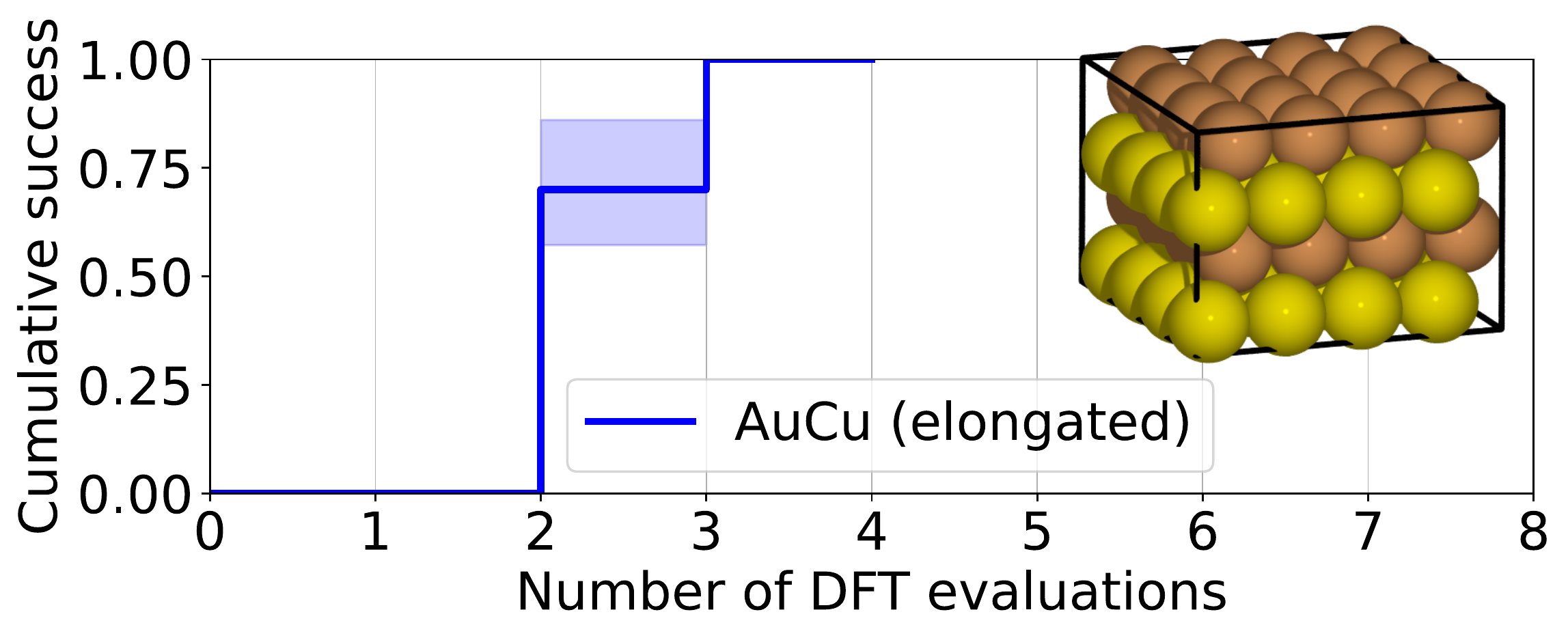}
    \caption{Success curve for optimizing the ordering of the atoms in AuCu bulk with an elongated fcc lattice using ICE-BEACON. A total of 10 individual optimization runs are performed. The number of DFT evaluations in the training set is shown on the $x$-axis, while the fraction of successful runs, where the correct structure is identified, is shown on the $y$-axis.}
    \label{fig:si_aucu}
\end{figure}

First, we optimize the unit cell and the lattice of AuCu bulk with the UnitCellFilter module in ASE package \cite{ase, unitcellfilter}. After the optimization, we see that the cubic symmetry of the lattice is broken and the Au and Cu layers are separated from each other more than in the fcc lattice. Using the obtained coordinates, we run 10 individual ICE-BEACON runs starting from random element fractions, and the resulting success curve is shown in Fig. \ref{fig:si_aucu}. 7 of the runs find the (correct) layered structure after 2 DFT calculations, and 3 of the runs find it after 3 DFT calculations.

~\\[20pt]
{\bf{CO@CuNi surface}}

\begin{figure}[h]
    \centering
    \includegraphics[width=1.0\linewidth]{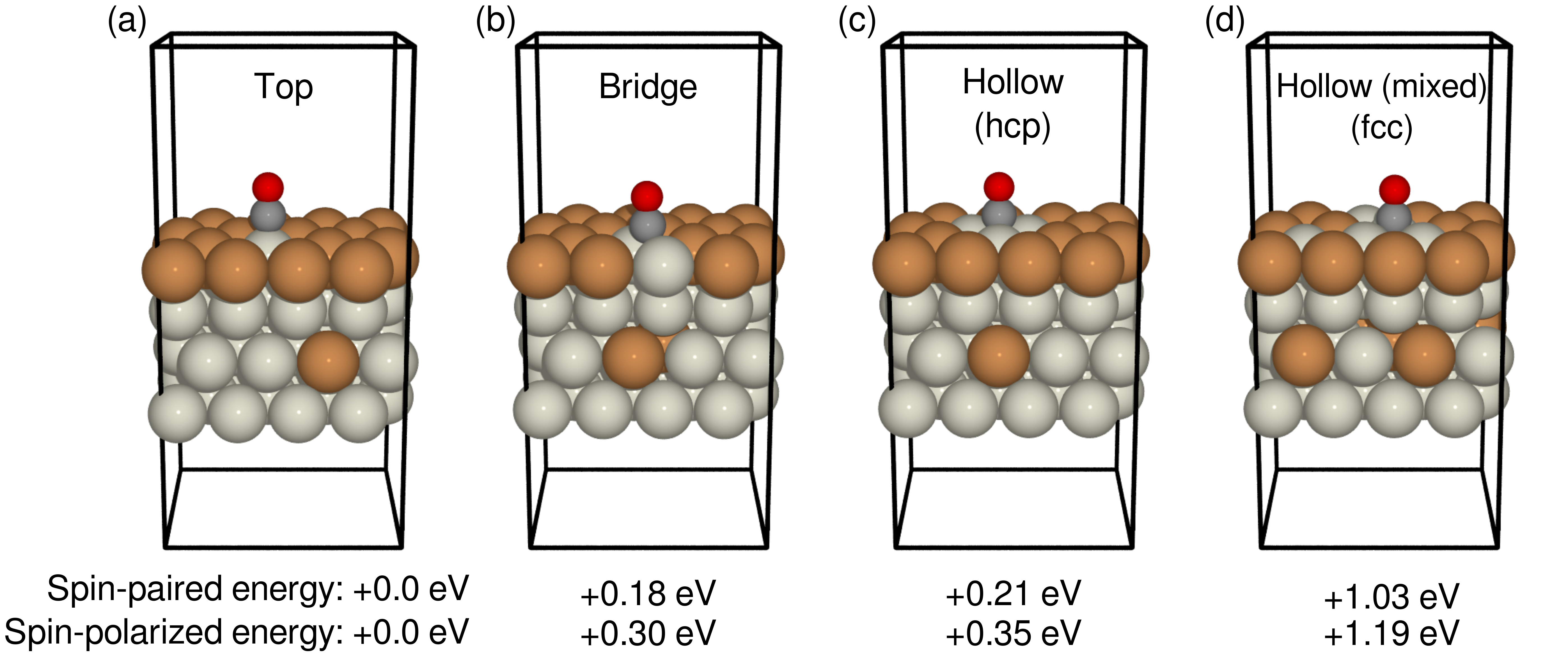}
    \caption{DFT energies for different CO@CuNi(111) systems. The text inside the unit cell indicates the adsorption site of the CO molecule. In (d), the word ``mixed'' refers to the surface layer that contains Ni atoms also in non-neighboring sites to CO. Below the structures, the energies are shown relative to the lowest energy structure separately for spin-paired and spin-polarized calculations.}
    \label{fig:si_surface}
\end{figure}

To verify that the spin polarization has no major effect on the energies of the final structures, we calculate the spin-polarized energy for the proposed global minimum structure from each of the 8 runs for {CO@CuNi}(111) surfaces. The resulting energies are shown in Fig. \ref{fig:si_surface}. We see that the energy ordering between the structures does not change if the spin-polarization is switched on. Thus we assume that the spin polarization caused by the narrow and partially occupied 3{\em{d}} band of Ni has no essential effect on the final results.




\printbibliography

\pagebreak

